\documentclass[pra,twocolumn,showpacs,superscriptaddress,tightenlines,
amsmath,amssymb,floatfix]{revtex4}
\usepackage{graphicx}
\newcommand{\beq}{\begin{equation}}
\newcommand{\eeq}{\end{equation}}
\newcommand{\bea}{\begin{eqnarray}}
\newcommand{\eea}{\end{eqnarray}}
\newcommand{\ba}{\begin{array}}
\newcommand{\ea}{\end{array}}
\newcommand{\bc}{\begin{center}}
\newcommand{\ec}{\end{center}}

\newcommand{\bml}{\begin{subequations}}
\newcommand{\eml}{\end{subequations}}
\newcommand{\commentout}[1]{{}}
\newcommand{\bk}{{\bf k}}

\newcommand{\adag}{a^\dagger}
\newcommand{\bdag}{b^\dagger}
\newcommand{\cdag}{c^\dagger}
\newcommand{\gdag}{g^\dagger}

\newcommand{\half}{\hbox{$\frac{1}{2}$}}
\newcommand{\fourth}{\hbox{$\frac{1}{4}$}}

\newcommand{\eq}[1]{(\ref{#1})}
\newcommand{\etal} {{\it et al.\/}}

\newcommand{\vol}[1]{{\bf #1}}
\newcommand{\comment}[1]{{}}
\begin{document}

\title{Feshbach-resonant  Raman photoassociation in a Bose-Einstein condensate}
\author{Matt Mackie}
\affiliation {Department of Physics, Temple University, Philadelphia, PA 19122}
\affiliation{Department of Physics, University of Connecticut, Storrs, CT 06268}
\author{Pierre Phou}
\affiliation {Department of Physics, Temple University, Philadelphia, PA 19122}
\author{Heather Boyce}
\affiliation {Department of Chemistry, Temple University, Philadelphia, PA 19122}
\author{Mannix Shinn}
\affiliation {Department of Physics, Temple University, Philadelphia, PA 19122}
\author{Lev Katz}
\affiliation {Department of Biology, Temple University, Philadelphia, PA 19122}

\date{\today}

\begin{abstract}
We model the formation of stable heteronuclear molecules via pulsed Raman photoassociation of a two-component Bose-Einstein condensate near a strong Feshbach resonance, for both counterintuitive and intuitive pulse sequencing. Compared to lasers alone, weak Raman photoassociation is enhanced by as much as a factor of ten (five) for a counterintuitive (intuitive) pulse sequence, whereas strong Raman photoassociation is barely enhanced at all--regardless of pulse sequence. Stronger intra-atom, molecule, or atom-molecule collisions lead to an expected decrease in conversion efficiency, but stronger ambient inter-atom collisions lead to an unexpected increase in the efficiency of stable molecule production. Numerical results agree reasonably with an analytical approximation.
\end{abstract}

\pacs{03.75.Fi,05.30.Fk,32.80.Wr}

\maketitle

\section{Introduction}

Not long after a Bose-Einstein condensate of atoms was created~\cite{AND95}, the race began to reach the molecular milestone, and thereby further enable fundamental studies~\cite{HUD06,SKO99}, practical applications~\cite{DEM02}, and proxy investigations~\cite{BUE07,NAT10}. The problem is that the laser cooling techniques that enable atomic condensation are difficult--though not impossible--to apply to molecules~\cite{SHU09}, while Stark deceleration~\cite{BET03} and buffer gas cooling~\cite{WEI98} have yet to reach quantum degeneracy. Alternatively, association of atoms into molecules--driven by a laser~\cite{THO87} or magnetic field~\cite{STW76}--is highly efficient at quantum degenerate phase space densities~\cite{BUR96}, and a fundamental coherence means that a condensate of atoms could be associated into a condensate of molecules~\cite{JAV99}. Indeed, high efficiency~\cite{WYN00,HOD05} and atom-molecule coherence~\cite{CLA02,WIN05} have been demonstrated in both photoassociation~\cite{WYN00,WIN05} and magnetoassociation~\cite{HOD05,CLA02}, and quantum degeneracy has been achieved on short time scales with magnetoassociation~\cite{KXU03}. However, production of a long-lived condensate of molecules in the absolute ground state is hampered by collisions between particles and limited laser intensity~\cite{DRU02}, and pre-loading an atomic condensate into an optical lattice~\cite{JAK98} was therefore proposed to mitigate collisional effects in association~\cite{JAK02}. So far, a quantum degenerate gas of molecule in the absolute ground state has been produced by several groups~ \cite{NI08}, but melting of the lattice to create a bulk condensate of stable molecules has yet to be achieved.

\begin{figure}
\centering
\includegraphics[width=8.5cm]{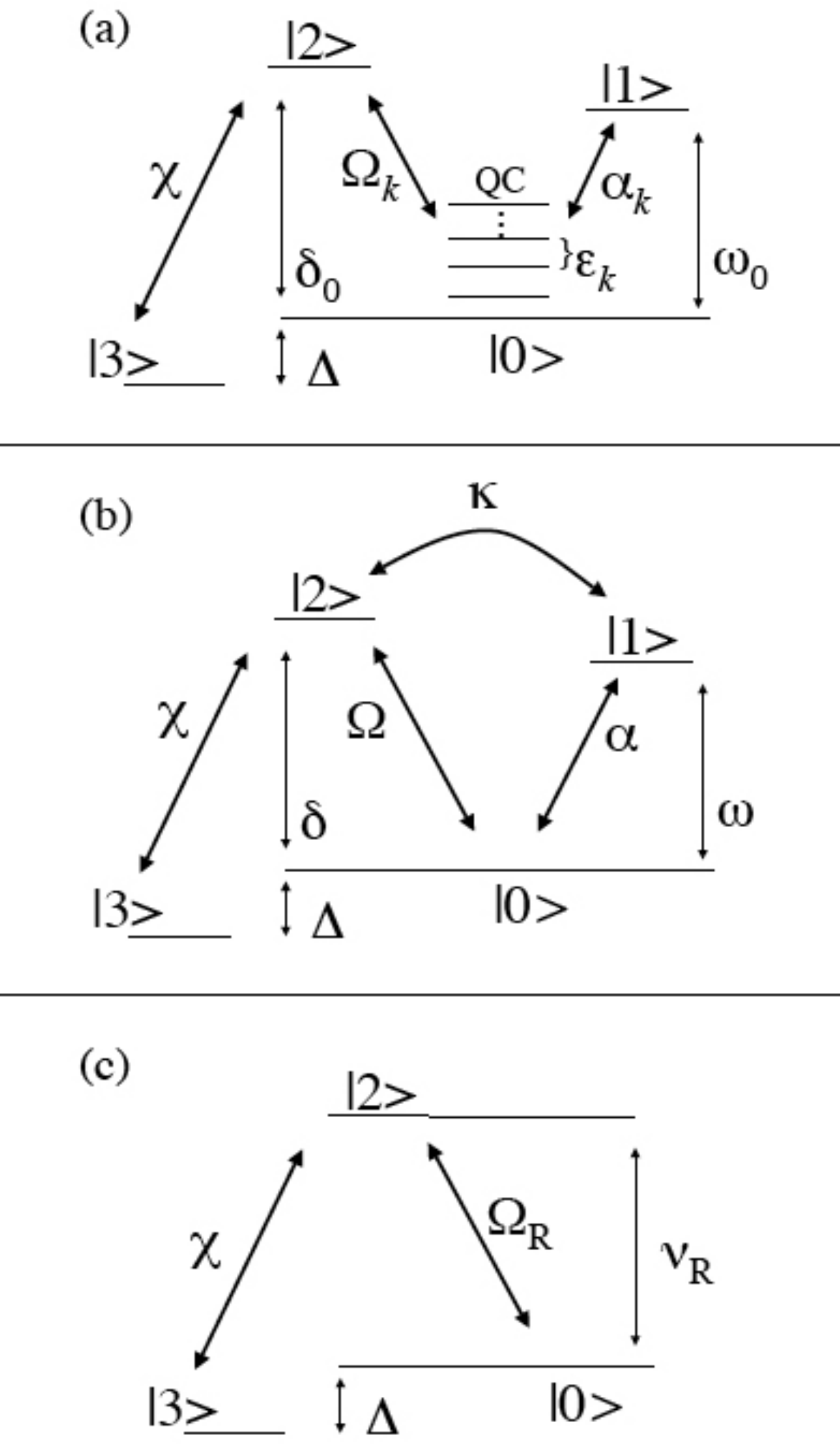}
\caption{
Few-level diagram of Raman photoassociation near a Feshbach resonance. In a full model (a), the Feshbach and photoassociation molecules share the same quasicontinuum (QC) of dissociation states. In a four-level approximation (b), the virtual quasicontinuum shifts the photoassociation and Feshbach detunings, and effectively couples transitions between the two excited molecular states. In the three-level approximation (c), the virtual Feshbach molecular state shifts the photoassociation detuning, and effectively renormalizes the photoassociation coupling and detuning.
}
\label{FEWL}
\end{figure}

Here we explore a different route to forming a bulk condensate of molecules in the absolute ground state. In particular, combining the photoassociation and Feshbach resonances has been shown to enhance the photoassociation rate constant~\cite{COU98,JUN08,MAC08} of a Bose-Einstein condensate by an order of magnitude~\cite{JUN08,MAC08}, due to constructive quantum interference between molecules formed by direct photoassociation and molecules formed by photoassociation via the Feshbach molecular state~\cite{MAC08}. We therefore consider bound-free-bound-ground transitions [Fig.~\ref{FEWL}(a)], whereby a magnetic field converts atom pairs into vibrationally-excited Feshbach molecules, a photoassociation pump laser converts atoms pairs into electronically-excited molecules, and a secondary dump laser converts the photoassociated molecules into stable molecules~\cite{SPECTROSCOPY,KOK01}. The lasers are pulsed to best avoid irreversible losses, and we consider counterintuitive sequencing where the dump pulse precedes the pump pulse, as well as intuitive sequencing where the pump pulse precedes the dump pulse. The model includes relevant elastic collisions between particles, dissociative decay of the Feshbach molecules, as well as spontaneous and dissociative decay of the electronically-excited photoassociation molecules. A quasicontinuum model explicitly includes dissociation and Feshbach molecules, whereas a resonant-interaction model treats the dissociation continuum and Feshbach molecules as virtual--leading to a magnetically-tunable photoassociation coupling and dissociation rate, in addition to the usual magnetically-tunable collisional interaction. Both the full and resonant-interaction models contrast with a previous model~\cite{MAC04} where the Feshbach resonance was accounted for merely with a magnetically tunable collisional interaction.

The question is whether Feshbach enhancement of primary photoassociation into excited molecules will carry over into enhancement of Raman photoassociation into stable molecules. On one hand hand, conversion in the counterintuitive scheme is presently hobbled by a combination of collisions and laser intensity limited by condensate size~\cite{DRU02}, and Feshbach-enhancement of the photoassociation coupling delivers strong coupling at low intensity which, in turn, could mitigate collisions. On the other hand, only an odd number of intermediate levels can form a dark state~\cite{ORE92}, and a counterintuitive scheme involving Feshbach and photoassociation molecules should therefore fall short of its trademark unit efficiency. Then again, it is possible that the dissociation continuum could act like a distinct intermediate level, as it has on a separate occasion~\cite{JAV02}, enabling a fully effective counterintuitive scheme. Finally, if a counterintuitive scheme is less than perfect, the question of enhancement over laser alone still remains and, moreover, whether or not it outperforms the intuitive scheme. All told, whereas strong photoassociation is already saturated and thus essentially un-enhanced, the Feshbach resonance enhances weak Raman photoassociation in a condensate and, somewhat surprisingly, the enhancement is stronger for stronger ambient inter-atomic collisions.

Our work is outlined as follows. In Sec.~\ref{MODEL}, we introduce the full model that includes elastic collisions between particles, a shared dissociation quasicontinuum for the Feshbach and photoassociation molecules, as well as spontaneous decay of the photoassociation molecules. Also, we develop the resonant-interaction model based on virtual continuum and Feshbach states, wherein the photoassociation interaction, the photodissociation rate, and the $s$-wave collisional interaction between atoms are magnetically tunable. Section~\ref{COMP} briefly reviews the computational algorithm and provides parameters for numerical experiments. Section~\ref{RESULTS} reports results for the efficiency of stable molecule production as a function of magnetic field, and compares these results against the two-photon rate constant. Finally, a summary is given in Section~\ref{SUM}. 

\section{Quasicontinuum and Resonant-Interaction Models}
\label{MODEL}

We focus on a two-component condensate because the resulting heteronuclear molecules are of significant recent interest for their dipolar properties~\cite{HUD06,SKO99,DEM02} (see also Refs.~\cite{HUT10}). Nevertheless, any dipolar interaction is considered relevant only after the molecules are formed, and is not accounted for in the association process. We also expect the results to apply to homonuclear systems.

In the few-level description illustrated in Fig.~\ref{FEWL}(a), we consider $N_1$ ($N_2$) atoms of species 1~(2) that have Bose condensed into the state $|0_1\rangle$ ($|0_2\rangle$), say, the plane-wave state with zero momentum $\hbar\bk=0$, which are represented in Fig.~\ref{FEWL}(a) as a single state $|0\rangle=|0_1\rangle|0_2\rangle$. A magnetic field tuned nearby a Feshbach resonance then couples two atoms in $|0\rangle$, one from each condensate, to a vibrationally-excited molecule in the state $|2\rangle$. Additionally, a photoassociation pump laser couples the same two atoms to an electronically-excited molecule in the state $|3\rangle$, and a secondary dump laser couples the molecule in $|3\rangle$ to a molecule in the absolute ground state $|4\rangle$. In the full model of Fig.~\ref{FEWL}(a), the Feshbach and photoassociation molecules dissociate into non-condensate atom pairs that occupy one of a quasi-continuum of states, say, plane-wave states of momentum $\pm\hbar\bk\neq0$.

In second-quantized notation, the Hamiltonian corresponding to Fig.~\ref{FEWL}(a) is 
\beq
H=H_M+H_P+H_D+H_C, 
\eeq
where the contribution due to the magnetic field is
\beq
\frac{H_M}{\hbar} = \omega_0 \cdag c 
  +\sum_\bk \alpha_\bk(\cdag a_{\bk,1}a_{-\bk,2} +\adag_{-\bk,2}\adag_{\bk,1}c),
\eeq
the contribution due to the photoassociation pump laser is
\beq
\frac{H_P}{\hbar} = \tilde\delta_0\bdag b 
  +\sum_\bk \Omega_\bk(\bdag a_{\bk,1}a_{-\bk,2} +\adag_{-\bk,2}\adag_{\bk,1}b),
\eeq
the contribution due to the secondary dump laser is
\beq
\frac{H_D}{\hbar} = -\Delta \gdag g 
  +\chi(\gdag b+\bdag g),
\eeq
and the contribution due to s-wave collisions in the condensates is
\bea
\frac{H_C}{\hbar} &=& \lambda_{00} \cdag\cdag cc + \lambda_{03} \cdag c\gdag g \nonumber\\&&
  +\half\sum_i\adag_ia_i\left(\lambda_{0i}\cdag c +\lambda_{3i}\gdag g
     +\sum_j \lambda_{ij}\adag_j a_j\right).
     \nonumber\\
\eea
Here atoms in the $i$th condensate are represented by $a_i=a_{i,0}$, atoms with momentum $\hbar\bk$ by $a_{i,\bk}$, Feshbach molecular condensate by $c_0=c$, photoassociation molecules by $b_0=b$, and stable molecular condensate by $g_0=g$. The detuning of the magnetic field from the Feshbach resonance is $\omega_0$, spontaneous decay of the Feshbach molecules~\cite{KOE05} is neglected, the one-photon laser detuning is $\delta_0=\Re[\tilde\delta_0]$, the spontaneous decay rate for the photoassociation molecule is $\Gamma_s=2\Im[\tilde\delta_0]$, and the two-photon detuning is $\Delta$. The magnetic-field coupling between the atoms and the Feshbach molecules is $\alpha_\bk=\alpha f_{M,\bk}$, the pump-laser coupling between the atoms and the photoassociation molecules is $\Omega_\bk=\Omega f_{P,\bk}$, and the dump-laser coupling between the photoassociation and stable molecules is $\chi$. The momentum dependence of the Feshbach and photoassociation couplings are contained in $f_{M,\bk}$ and $f_{P,\bk}$, respectively, where $f_{\bk=0}=0$. Finally, the strength of collisions is determined by $\lambda_{ij}$, which is determined by the s-wave scattering length. Compared to the spontaneous decay rate, elastic collision involving primary photoassociation molecules are neglected, and we also neglect vibrational relaxation~\cite{BAL98} of the photoassociation and Feshbach molecules.

The quasicontinuum mean-field model is derived from a $c$-number approximation to the Heisenberg equations, $i\hbar\dot{x}=[x,H]$, with $x$ is the relevant operator, which generally works best for $N\agt100$~\cite{KOS00}. Dissociation of Feshbach and photoassociation molecules into non-condensate atoms pairs is accounted for with the operator $a_{\bk,1}a_{-\bk,2}$, and the corresponding $c$-number amplitude $A_\bk=\langle a_{\bk,1}a_{-\bk,2}\rangle$. Lastly, the quasicontinuum in momentum is converted into a continuum in frequency according to
$\sum_\bk \rightarrow N/(4\pi^2\omega_\rho^{3/2})\int d\epsilon$, where $\hbar\epsilon=\hbar^2 k^2/(2\mu)$ is the kinetic energy and $\omega_\rho=\hbar\rho^{2/3}/(2\mu)$ is the characteristic frequency for a dissociated pair, with $\rho$ the total particle density and $\mu$ the reduced atomic mass. The resulting equations of motion are given by
\bml
\bea
i\dot{a_1} &=& \Lambda_1a_1 +\alpha a_2^*c +\Omega a_2^*b,\\
i\dot{a_2} &=& \Lambda_2a_2 +\alpha a_1^*c +\Omega a_1^*b,\\
i\dot{A}(\epsilon) &=& \epsilon A(\epsilon) +\alpha f_M(\epsilon)c 
  +\Omega f_P(\epsilon)b,\\
i\dot{c} &=&(\omega_0+\Lambda_c) c +\alpha a_1a_2 
  +\xi_M\int d\epsilon f_MA,\\
i\dot{b} &=& \tilde\delta_0b +\Omega a_1a_2 +\chi g +\xi_P\int d\epsilon f_PA,\\
i\dot{g} &=& -(\Delta -\Lambda_g)g +\chi b.
\eea
\label{FULL_EQM}
\eml
Defining $\Lambda_{i\neq j}=\half\rho\lambda_{ij}$ and $\Lambda_{i=j}=\rho\lambda_{ij}$, the respective mean-field shifts are $\Lambda_1=\Lambda_{11}|a_1|^2+\Lambda_{12}|a_2|^2$, $\Lambda_2=\Lambda_{22}|a_2|^2+\Lambda_{12}|a_1|^2$, $\Lambda_c=\Lambda_{00}|c|^2+\Lambda_{03}|g|^2$, and $\Lambda_g=\Lambda_{33}|g_2|^2+\Lambda_{31}|a_1|^2 + \Lambda_{32}|a_2|^2 + \Lambda_{30}|c|^2$.  Also, $\xi_M=\alpha/(4\pi^2\omega_\rho^{3/2})$ is the magnetodissociation coupling and $\xi_P=\Omega/(4\pi^2\omega_\rho^{3/2})$ is the photodissociation coupling.

The resonant-interaction model is derived by first treating the dissociated pair amplitude adiabatically ($\dot{A}=0$), which is equivalent to the limit of weakly-bound molecules~\cite{MAC08}, and leads to the effective four-level system~[Fig.~\ref{FEWL}(b)] with mean-field equations of motion
\bml
\bea
i\dot{a_1} &=& \Lambda_1a_1 +\alpha a_2^*c +\Omega a_2^*b\\
i\dot{a_2} &=& \Lambda_2a_2 +\alpha a_1^*c +\Omega b a_1^*b\\
i\dot{c} &=&(\tilde\omega+\Lambda_c)c +\alpha a_1a_2 +\kappa b \\
i\dot{b} &=& \tilde\delta b +\Omega a_1a_2 +\chi g +\kappa c\\
i\dot{g} &=& -(\Delta-\Lambda_g)g +\chi b.
\eea
\label{4L_EQM}
\eml
The virtual continuum leads to an effective coupling between the Feshbach and photoassociation molecules~\cite{MAC08} of strength
\beq
\kappa=\frac{1}{4\pi}\,
 \frac{\alpha\Omega}{\omega_\rho^{3/2}}\,
   \Re\left[\lim_{\epsilon_0\rightarrow0}\int d\epsilon
     \sqrt{\epsilon}\frac{f_M(\epsilon)f_P(\epsilon)}{(\varepsilon-\epsilon_0)}\right].
\eeq
Additionally, there are real and imaginary shifts for each molecular detuning: $\tilde\omega=\omega_0-\sigma_m-i\gamma_M/2$ and $\tilde\delta=\tilde\delta_0-\sigma_p-i\gamma_P/2$, where $\sigma_{M(P)}=\Re[\Sigma_{M(P)}]$ and $\gamma_i=\Im[\Sigma_{M(P)}]$ with
\bml
\bea
\Sigma_M &=& \fourth\alpha\xi_M
  \left[\lim_{\epsilon_0\rightarrow0}
    \int d\epsilon\sqrt{\epsilon}\frac{f_M^2(\epsilon)}{(\epsilon-\epsilon_0)}\right],
\\
\Sigma_P &=& \fourth\Omega\xi_P
  \left[\lim_{\epsilon_0\rightarrow0}
    \int d\epsilon\sqrt{\epsilon}\frac{f_P^2(\epsilon)}{(\epsilon-\epsilon_0)}\right].
\eea
\eml
The real shift is the well known result of coupling a bound state to a continuum~\cite{FED96}, and the imaginary shift is the dissociation rate. Of course, here the Feshbach shift $\sigma_M$ is static, while the photoassociation $\sigma_P$ shift is transient, and both are treated as implicit in the detuning rather than explicitly. Next, we arrive at the resonant-interaction model by treating the Feshbach amplitude adiabatically, which is the limit of large detuning of the magnetic field from the Feshbach resonance, and leads to an effective three-level system,
\bml
\bea
i\dot{a_1} &=& \Lambda_1'a_1 +\Omega_R a_2^*b\\
i\dot{a_2} &=& \Lambda_2' a_2 +\Omega_R a_1^*b\\
i\dot{b} &=& \tilde\nu b +\Omega_R a_1a_2 +\chi g\\
i\dot{g} &=& -(\Delta-\Lambda_g) g +\chi b.
\eea
\label{3L_EQM}
\eml

As detailed previously~\cite{COU98,JUN08,MAC08}, in the resonant-interaction model the Feshbach resonance effectively modifies the photoassociation interaction $\Omega_R=\Omega-\alpha\kappa/\omega$ and detuning $\tilde\nu=\nu_R-i\Gamma_R/2$, where the effective detuning is $\nu_R=\delta-\kappa^2/\omega$ and the effective decay rate is $\Gamma_R=\Gamma+(\kappa^2/\omega^2)\gamma_M$ with $\Gamma=\Gamma_s+\gamma_P$. Depending on the sign of the Feshbach detuning, the modified photoassociation coupling can be greater than the unmodified coupling, zero, or less than the unmodified coupling. Similarly, the real part of the resonant contribution to the photoassociation detuning, $\kappa^2/\omega$, can produce a redshift, no shift, or a blueshift, as opposed to the ambient shift, $\sigma_P$, which is strictly to the red~\cite{FED96}. Moreover, the resonant contribution to the decay rate, $(\kappa^2/\omega^2)\gamma_M$, leads to decay that depends on magnetic field and diverges near the Feshbach resonance. Finally, we find the usual resonant collisional interaction, $\Lambda_1'=\Lambda_{11}|a_1|^2+\Lambda_R|a_2|^2$ and $\Lambda_2'=\Lambda_{22}|a_2|^2+\Lambda_R|a_1|^2$, where $\Lambda_R=\Lambda_{12}-\alpha^2/2\omega$. In one-photon transitions~\cite{MAC08}, collisions--resonant or otherwise--are neglected compared to the spontaneous decay rate, but here the timescale for conversion is long enough, especially in the counterintuitive scheme, that collisions become relevant.

\section{Numerical Details}
\label{COMP}

For either the quasicontinuum model~\eq{FULL_EQM} or resonant-interaction model~\eq{3L_EQM}, the mean-field equations of motion are written in matrix form, $i\hbar\dot\psi=M\psi$, where the nonlinear matrix $M(\psi)$ depends on the wavevector $\psi$. Given the solution $\psi(t)$, the solution $\psi(t+dt)$ is then approximated as $\psi(t+dt)=M_-M_+^{-1}\psi(t)$, where $M_\pm=1\pm i(dt/2)M$. Given the solution to $M_+\phi(t)=\psi(t)$, the desired solution is simply $\psi(t+dt)=M_-\phi(t)$. The nonlinearity is accounted for with a two-step predictor-corrector method, wherein the prediction is $\psi_p(t+dt)=M_-\phi(t)$, and the correction is then $\psi(t+dt)=\bar{M}_-\bar\phi(t)$ with $\bar\phi(t)=\bar{M}_+\psi(t)$, $\bar{M}_\pm=M_\pm(\bar\psi)$ and $\bar\psi(t)=[\psi_p(t)+\psi(t)]/2$. This algorithm has the advantage that it can be made to scale linearly with the number of states~\cite{MAC04,JAV02}, which is important in the quasicontinuum model since we fix the number of quasicontinuum states to $10^6$.

Considering the parameter values, the strength of the free-bound association couplings are measured relative to the characteristic frequency $\omega_\rho$~\cite{JAV02,KOS00} and, for consistency, so are the collisional couplings. Hence, we consider a strong Feshbach coupling $\alpha\gg\omega_\rho$, as well as both strong ($\Omega_0\gg\omega_\rho$) and weak ($\Omega_0\alt\omega_\rho$) photoassociation, respectively. The laser pulses are modeled as Gaussian, $\Omega=\Omega_0\exp[-(t-T_1)^2/\tau^2]$ and $\chi=\chi_0\exp[-(t-T_2)^2/\tau^2]$, so that $\kappa=\kappa_0\exp[-[(t-T_1)^2/\tau^2]$. Finally, collisions are pre-mitigated to a certain extent with low density~\cite{MAC04} $\rho=10^{12}$~cm$^{-3}$, and we assign elastic collisional couplings that range from weak to moderate.

In particular, the Feshbach coupling is $\alpha=134.8$, the weak (strong) photoassociation pump coupling is $\Omega_0=15.4$ (154), and the spontaneous decay rate is $\Gamma_s=41\Omega_0$. Here strong photoassociation corresponds to a laser set at the saturation intensity and, since $\Omega\propto\sqrt{I}$, the weak coupling corresponds to an intensity two orders of magnitude below the saturation intensity~\cite{SAT_NOTE,BOH99,MAC10,SCH02}. In the counterintuitive scheme, the dump coupling is $\chi_0=50\Omega_0$, but the intuitive scheme is more efficient for $\chi_0=\Omega_0$. The respective coupling to the dissociation continuum is then determined by the Lorentzian $f_i=1/(1+\varepsilon^2/\beta_i^2)$, which is in turn determined by the molecular size $\beta_i=\hbar/mL_i^2$. For magnetoassociation we choose a point particle, $L_M=\rho^{-1/3}$, and for photoassociation we choose a typical size, $L_P=130a_0$, where $a_0$ is the Bohr radius. For the collisional couplings, we choose values for $\lambda_{ij}$ such that
$\Lambda_{00}=\Lambda_{30}=0.8$,
$\Lambda_{11}=5.1\times10^{-3}$, 
$\Lambda_{22}=2.1\times10^{-2}$, 
$\Lambda_{12}=4\times10^{-2}$,
$\Lambda_{33}=8.1\times10^{-3}$, 
$\Lambda_{31}=4.3\times10^{-2}$, and 
$\Lambda_{32}=1.9\times10^{-2}$. 
Finally, in the tunable collision model, the magnetodissociation rate is $\gamma_M/\Gamma_s=7.4\times10^{-2}$, the photodissociation rate for strong (weak) coupling is~\cite{SAT_NOTE} $\gamma_P/\Gamma_s=1$ ($1/100$), and the peak cross-molecular coupling is $\kappa_0=3.9\times10^3$ ($3.9\times10^4$) for weak (strong) photoassociation. 

To give the counterintuitive scheme the best chance for success, i.e., slow enough to foster adiabaticity but fast enough to outrun ambient collisions, we set the pulse width according to $\Omega_0\tau=5\times10^3$. Also, to ease the numerical overhead we only optimize the one-photon detuning $\delta$, but we fix the pulse delay to $D=T_1-T_2=-2\tau$, and we also fix the two-photon detuning to resonance ($\Delta=0$) in the weak case and Stark-shifted resonance ($\Delta=\Omega_0^2/2\delta$) in the strong case. For the intuitive scheme, coincident pulses ($D=0$) are generally optimal for all magnetic fields, both the pulse width and one-photon detuning are optimized at each magnetic fields, but the two-photon detuning is again fixed to resonance in the weak case and Stark-shifted resonance in the strong case. Although we do not go into details, for both pulse schemes the optimized one-photon detuning is consistent with the expected~\cite{MAC08,JUN08} dispersive-like behavior.

\section{Feshbach Enhancement}
\label{RESULTS}

\begin{figure}
\centering
\includegraphics[width=8.5cm]{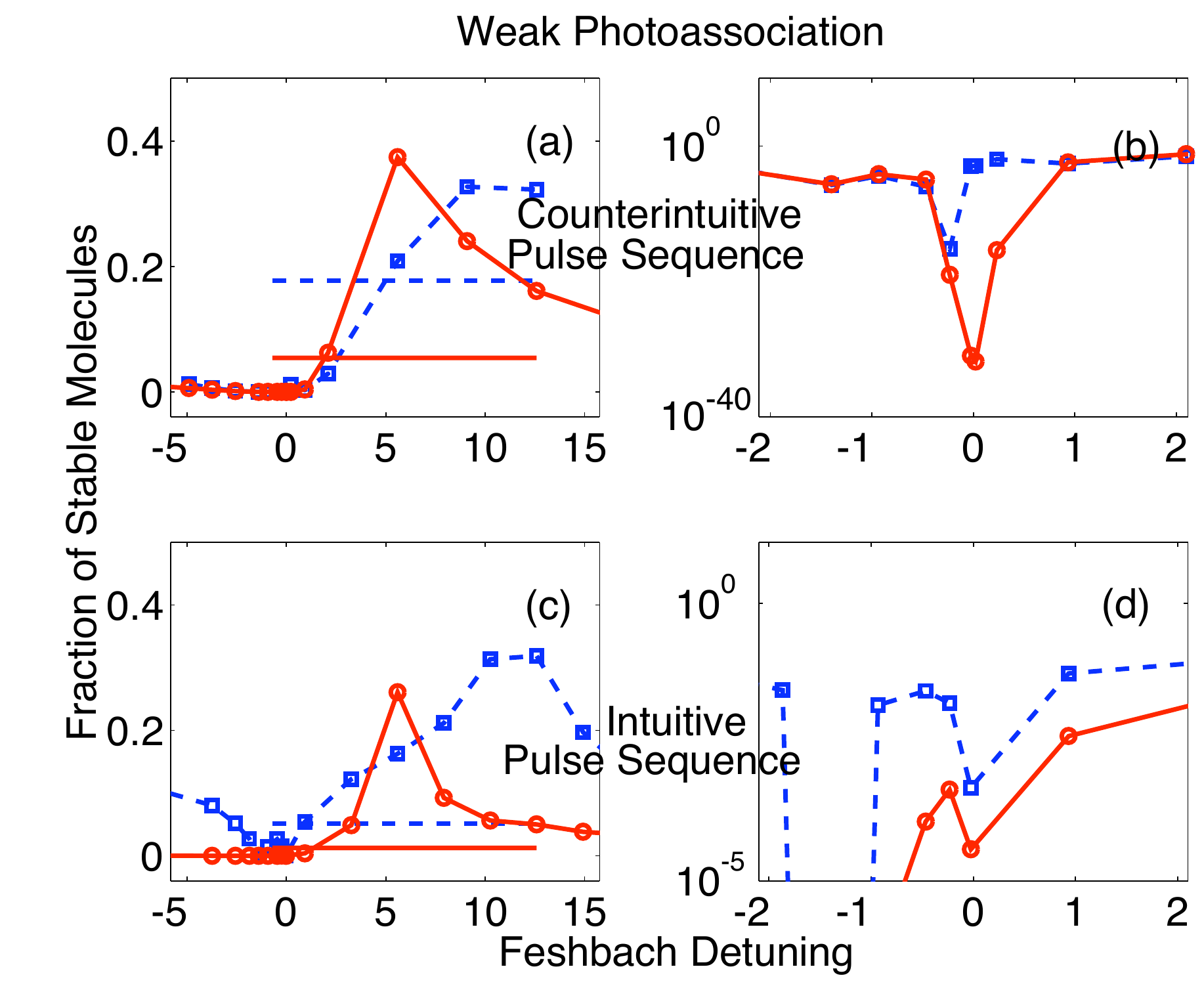}
\includegraphics[width=8.5cm]{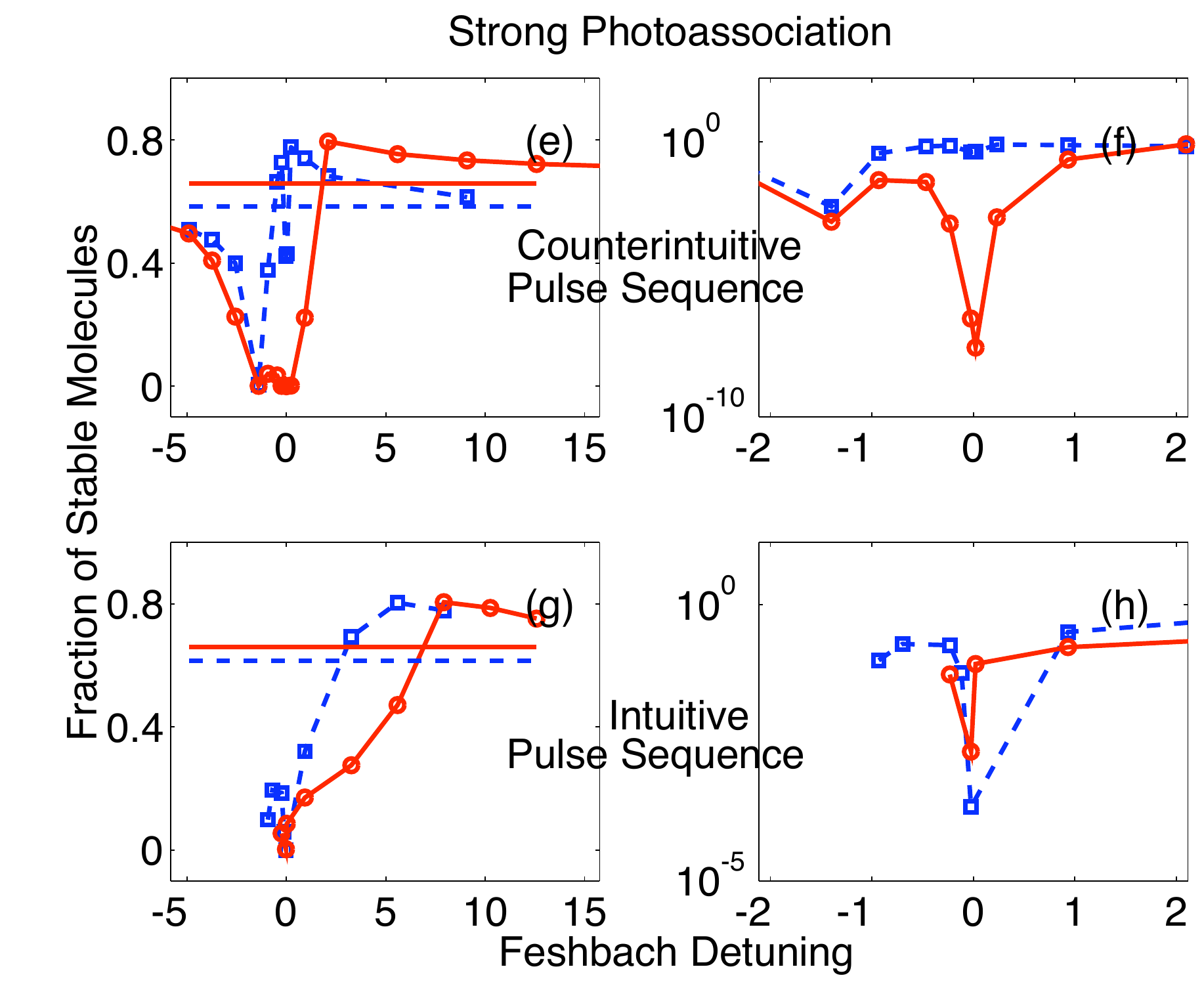}
\caption{(Color online) Feshbach-stimulated Raman photoassociation of a two-component Bose-Einstein condensate for weak (a-d) and strong photoassociation (e-h), for the magnetically-tunable (blue dashed,~$\Box$) and the full quasicontinuum (red solid, $\circ$) models. The dashed (solid) horizontal line is the quasicontinuum (tunable-collisions) result for zero magnetic field. The dimensionless Feshbach detuning is $\omega/\Gamma_s$.
}
\label{FULL_FIGURE}
\end{figure}

This section reports the results of numerical experiments, summarized in Fig.~\ref{FULL_FIGURE}.  The node below resonance arises from destructive interference between direct photoassociation and indirect photoassociation occurring via the Feshbach molecular state~\cite{MAC08}. A second node appears on resonance, and in the full model this node is due to the absence of an atom-molecule dark state that includes both excited molecules, while in the effective model is due to divergence of the magnetically-tunable photodissociation term. Regarding the magnitude of peak enhancement, for weak coupling the magnetic field enhances both schemes by about a factor of five, and the counterintuitive scheme outperforms the intuitive scheme by about $10\%$ in the full model, and the two are roughly tied in the resonant-interaction model. Strong photoassociation is enhanced very little, to about about 80\%, regardless of pulse order. The agreement with the full quasicontinuum model is reasonable given the simplicity of the effective model. It is worth noting that a model where the primary photoassociation molecule is virtual instead of the Feshbach molecule improves the agreement between the two models near magnetic resonance, but worsens the off-resonant disagreement.

Focusing on weak photoassociation, improvements to the counterintuitive scheme can be made by increasing the pulse area from $\Omega_0\tau=5\times10^5$ to $2.2\times10^5$ ($6\times10^5$) in the resonant-interaction (quasicontinuum) model, whereby the maximally-enhanced counterintuitive efficiency improves from 30\% to nearly 50\% (76\%). In the resonant-interaction model, further improvements can also be made by setting the pulse area according to $\Omega_R$ instead of $\Omega$, but this would require a change in the dump coupling to $\chi=50\Omega_R$, in order to satisfy the conditions for adiabatic following (specifically, $\chi\gg\Omega$ for $t\rightarrow-\infty$ and $\Omega\gg\chi$ for $t\rightarrow\infty$), and therefore would require an impractical amount of dump laser intensity as $\Omega_R$ diverges on resonance. More importantly, this improvement is misleading since a larger peak dump pulse does not improve conversion in the full quasicontinuum model, which we attribute to the absence of a dark state that includes both photoassociation and Feshbach molecules.

The dip near resonance and the off-resonant peak together indicate that Feshbach enhancement does not tell the whole story, since Feshbach enhancement peaks near magnetic resonance~\cite{MAC08,JUN08}. In the full model, cross-coupling between the photoassociation and Feshbach molecules arises due to the shared dissociation continuum~\cite{MAC04}, which effectively enhances the weak photoassociation coupling to be comparable to the Feshbach coupling, but it also enhances the Feshbach losses to be comparable to the photoassociation losses. In other words, the Feshbach molecular state decays vicariously through the photoassociation state, and peak enhancement therefore occurs where the Feshbach detuning is large compared to the spontaneous decay rate of the photoassociation state, roughly $\omega/\Gamma_s\approx5$ in Fig.~\ref{FULL_FIGURE}. Although a completely dark state is absent, it so happens that the photoassociation state is dark and the Feshbach state is dim. In weak photoassociation, collisions disrupt the dark state and conversion is independent of pulse sequence for short pulses, and longer pulses help the counterintuitive scheme until the timescale for collisions is reached. In strong photoassociation, collisions play a lesser role. For a counterintuitive pulse sequence, the dark photoassociation state then allows efficient conversion closer to the Feshbach resonance, and the enhancement peak shifts to the red. For an intuitive pulse sequence where the dark state is moot, stronger laser coupling requires larger Feshbach detuning to combat vicarious losses, and the peak shifts to the blue. As with lasers alone, the peak efficiency decreases with increasing strength of intra-atomic, molecular, or atom-molecule collisions.

In terms of the effective model, peak improvement in the weak case occurs roughly where the resonant collisional interaction vanishes $\Lambda_R=0$, or $\omega/\Gamma_s=\alpha^2/(4\Lambda_{12}\Gamma_s)\approx 9$, regardless of pulse sequence. For strong photoassociation, collisions again play a lesser role, and the counterintuitive scheme is more efficient closer to resonance, even closer than in the quasicontinuum model due to the faux dark state in the effective three-level model. For the intuitive pulse sequence, the difference is made up--perhaps coincidentally--by the peak laser contribution to the resonant collisional interaction, $-\Omega_0^2/(4\delta)$. For weak or strong photoassociation, with tunable collisions nulled, the conversion efficiency is determined by the non-tunable collisional interactions. That peak improvement occurs where the magnetically-tunable collisional interaction vanishes, and that the magnitude of peak improvement is determined by non-tunable collisions, is in line with previous work on Raman photoassociation of an interacting Bose condensate~\cite{DRU02,MAC04} combined with a far-detuned Feshbach resonance~\cite{MAC04}. So far, so good.

\begin{figure}
\centering
\includegraphics[width=8.5cm]{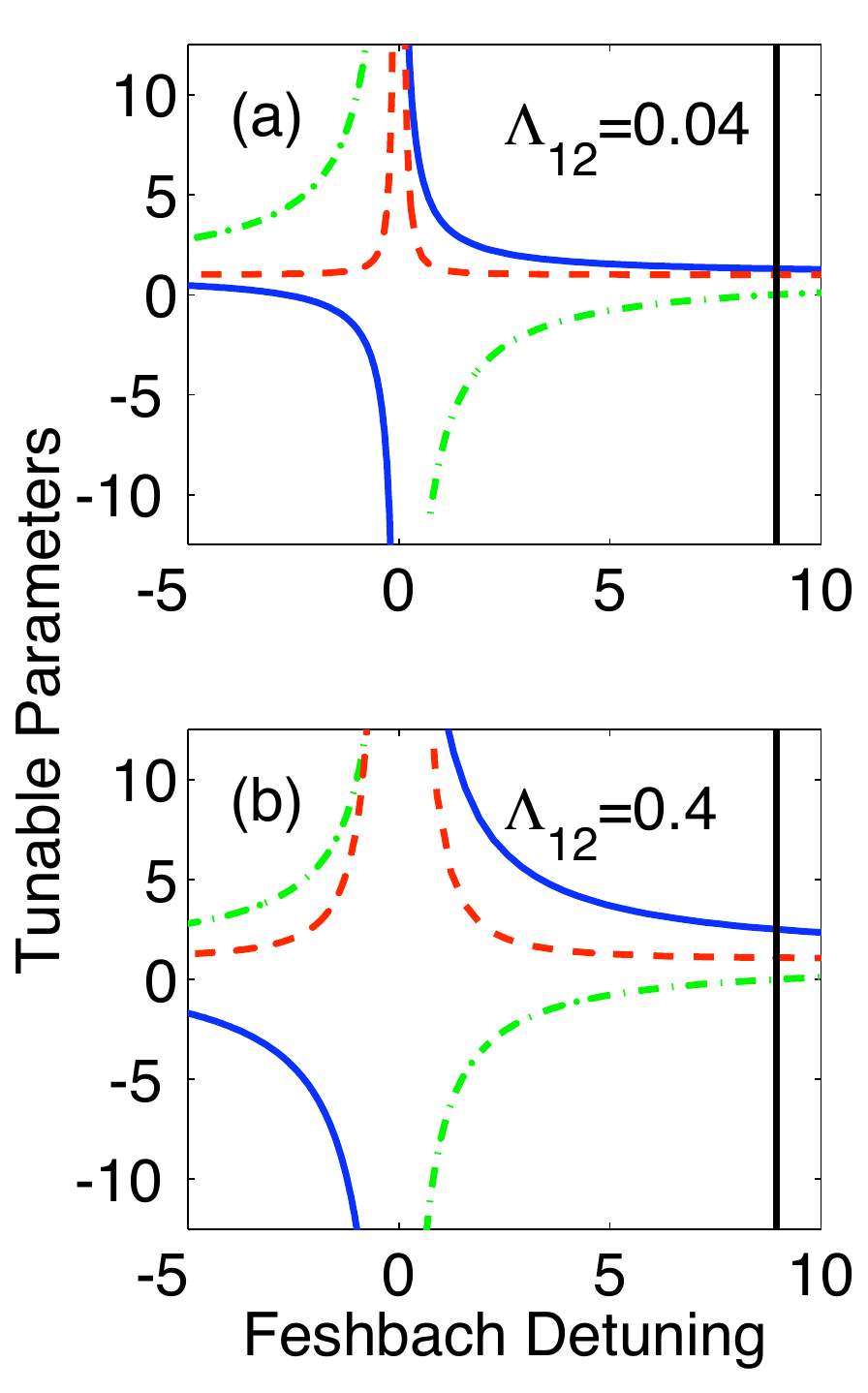}
\caption{(Color online) Magnetically-tunable parameters vs. Feshbach detuning for weak photoassociation, where the solid blue line is the photoassociation coupling ($\Omega_R/\Omega$), the dashed red line is the tunable decay rate for the photoassociation molecule ($\Gamma_R/\Gamma$), and the dot-dashed green line is the tunable collisional interaction ($\Lambda_T/\Lambda_{12}$). The vertical line denotes $\Lambda_R=0$,  the dimensionless Feshbach detuning is defined as $\omega/\Gamma_s$, and the parameters in panel (a) are the same as in Fig.~\ref{FULL_FIGURE}.
}
\label{TUNABLE}
\end{figure}

The role of Feshbach enhancement can be understood further by considering the magnetically tunable parameters in the effective model [Fig.\ref{TUNABLE}], where panel (a) corresponds to the parameters in Fig.~\ref{FULL_FIGURE}. The resonant decay rate drops relatively quickly to its ambient value, and the photoassociation coupling follows shortly thereafter, but it is not until the resonant collisional interaction reaches a reasonable value ($\omega/\Gamma_s\alt9$) that any improvement kicks in. At peak enhancement, the photoassociation coupling, weak or strong, is enhanced to $\Omega_R/\Omega=1.2$, which amounts to an effective increase in intensity of a factor of about~1.4. Part of the reason for the lackluster improvement in strong photoassociation is due to it already being saturated~\cite{FED96,KOS00,JAV02,MAC04}, so any increase in coupling is moot.

\begin{figure}
\centering
\includegraphics[width=8.5cm]{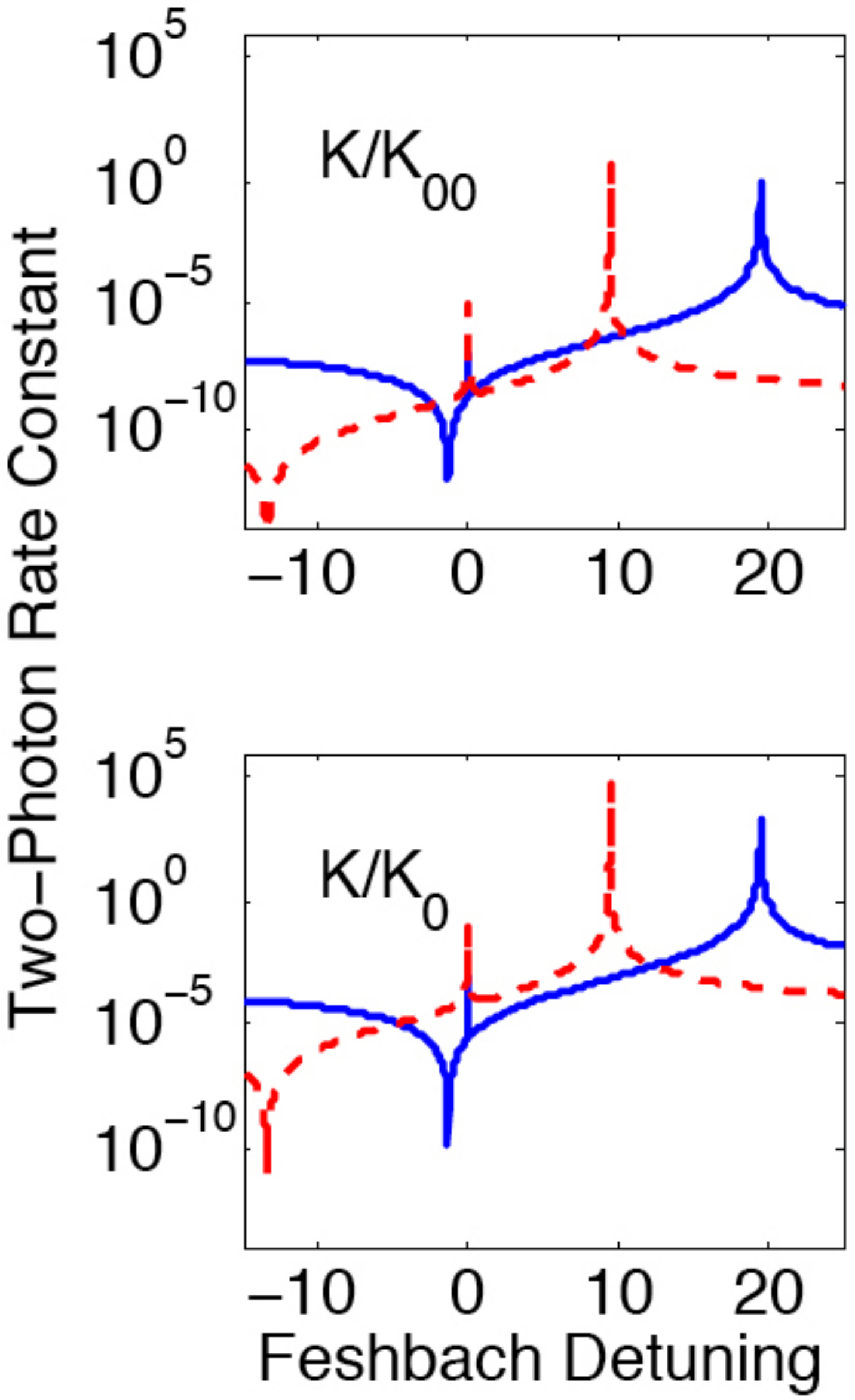}
\caption{(Color online) Feshbach enhanced two-photon rate constant for weak photoassocaition, where the solid (dashed) lines corresponds to $\Lambda_{12}=0.04$ (0.4). Here the dimensionless Feshbach detuning is again defined as $\omega/\Gamma_s$, and the dimensionless rate constant is defined relative to the un-enhanced two-photon rate constant for either a non-interacting ($K_{00}$) or an interacting ($K_0$) condensate (see text). 
}
\label{RATECON}
\end{figure}

The truly unexpected find is that, whereas an increase in the strength of non-tunable collisions (intra-atom, molecule, or atom-molecule) will decrease the conversion efficiency~\cite{DRU02,MAC04}, an increase in the ambient value of the tunable inter-atomic collisional interaction, $\Lambda_{12}$, will actually {\em increase} the efficiency of stable molecule production. In particular, since the Feshbach coupling~\cite{STW76} $\alpha\propto\sqrt{\Lambda_{12}}$, the Feshbach-detuning location of $\Lambda_R=0$, i.e., $\omega_z=\alpha^2/(4\Lambda_{12})$, is independent of $\Lambda_{12}$.  At the magnetic-field location of peak enhancement, $\omega_z$, the resonant contribution to the photoassociation coupling is $\alpha\kappa/(2\omega_z)\propto\Lambda_{12}$, and stronger inter-atomic collisions therefore lead to stronger Feshbach enhancement at $\omega=\omega_z$, as illustrated in Fig.~\ref{TUNABLE}(b) for weak photoassociation and $\Lambda_{12}\rightarrow10\Lambda_{12}$. For the counterintuitive scheme at a pulse area of $\Omega_0\tau=5\times10^3$, in the resonant-interaction model a factor of two (ten) increase in $\Lambda_{12}$ enhances conversion from 30\% to 37\% (60\%), and in the full quasicontinuum model the former (latter) increase in $\Lambda_{12}$ enhances conversion from 30\% to 42\% (67\%).

These numerical results are supplemented analytically as follows. Deriving a two-level system from Eqs.~\eq{3L_EQM} in the limit of large Feshbach-shifted detuning ($\nu\gg\Gamma_s$), and then deriving a rate equation for the atom losses~\cite{KOS00,MAC08}, we obtain the rate constant for Feshbach-enhanced Raman photoassociation for CW lasers
\beq
\rho K = \frac{1}{4}\,\frac{\chi_2^2\Gamma_2}{\sigma_{\rm mf})_R^2+\Gamma_2^2/4}\,,
\eeq
where the tunable two-photon coupling is $2\chi_2=\Omega_R\chi/\nu$, the tunable two-photon decay rate is $4\Gamma_2=(\chi/\nu)^2\Gamma_R$, and the tunable mean-field shift is approximated to the static value $\sigma_{\rm mf})_R=\sigma_{\rm mf}+2\Lambda_R$ with the non-resonant mean field shift $\sigma_{\rm mf}=\Lambda_{11}+\Lambda_{22}-(\Lambda_{13}+\Lambda_{23}+\Lambda_{33})$. Results shown in Fig.~\ref{RATECON} for weak photoassociation with $\nu=10\Gamma_s$ are broadly consistent with Fig.~\ref{FULL_FIGURE}. In particular, the dip below the Feshbach resonance corresponds to $\Omega_R=0$. The peak in atom losses at the Feshbach resonance corresponds to a peak in two-photon losses, which corresponds to the decreased molecule formation near Feshbach resonance in Fig.~\ref{FULL_FIGURE}. The peak far-above the Feshbach resonance arises from $\Lambda_R=\sigma_{\rm mf}/2$, which corresponds to the numerical peak for $\Lambda_R=0$. We attribute the difference to the static mean-field shift approximation that neglects transient populations, which also leads to a peak location that is not independent of $\Lambda_{12}$.  Nevertheless, for $\Lambda_{12}=0.04$ the enhanced two-photon rate constant  peaks at roughly the value for a non-interacting gas, $K_{00}=\chi_{20}^2/\Gamma_{20}$, and for $\Lambda_{12}=0.4$ it increases to $K\sim5K_{00}$. Moreover, compared to the $\alpha=0$ result for an interacting condensate, $4K_0=\chi_{20}^2\Gamma_{20}/[\sigma_{\rm mf)_0}+\Gamma_{20}^2/4]$, the Feshbach resonance enhances the rate two-photon constant by roughly three orders of magnitude for $\Lambda_{12}=0.04$, which increases to well over four orders of magnitude for $\Lambda_{12}=0.4$. Note that $2\chi_{20}=\Omega\chi/\delta$, $4\Gamma_{20}=(\chi/\delta)^2(\Gamma_s+\gamma_{\rm PA})$, and $\sigma_{\rm mf)_0}=\sigma_{\rm mf}+2\Lambda_{12}$.

Before closing, we emphasize that the two-photon detuning has not been optimized, and that the laser detunings in general have not been chirped~\cite{LIN04}, both of which could lead to further improvements. Also, while comparisons between thermal and condensate systems should be taken with a grain of salt, the results for Feshbach enhancement in a thermal gas~\cite{HAN11} indicate reduced efficiency upon averaging over density, and improved efficiency for narrower Feshbach resonance. Off hand, in association of an interacting condensate the atom-molecule coupling $\propto\sqrt{\rho}$ and the collisional coupling $\propto\rho$. Collisions therefore play the biggest role at the center of the trap, and inhomogeneity should have little effect on final conversion efficiencies in a local density approximation. Nevertheless, we look forward to a full investigation--including an explicit trapping potential--of condensate inhomogeneity for both wide and narrow Feshbach resonances.

\section{Summary}
\label{SUM}

In short, we find that a strong Feshbach resonance can substantially improve weak--but not strong--Raman photoassociation, independent of pulse ordering. The lack of Feshbach-enhancement in strong photoassociation is attributed to an already-saturated transition from atoms to molecules, and the independence of pulse ordering to an un-optimized pulse length for the counterintuitive pulse order. For weak and strong photoassociation, counterintuitive pulse sequences are indeed more efficient for larger pulse areas since the photoassociation molecular state is still dark (numerically), even if the Feshbach molecular state is dim.

In the quasicontinuum model, vicarious photoassociation losses from the Feshbach state mean that peak enhancement occurs when the Feshbach detuning is large compared to the photoassociation linewidth. In the resonant-interaction model, peak enhancement occurs where the resonant inter-atomic interaction vanishes. Also, disagreement between the resonant-interaction and quasicontinuum models on final conversion efficiencies and the nature of the dark state highlights the importance of explicitly including the Feshbach molecular state in modeling magnetoassociation.

Finally, whereas the peak conversion efficiency decreases for stronger intra-atomic, molecular and atom-molecule collisions, we find that the peak conversion efficiency actually increases for stronger inter-atomic collisions. Systems with a combination of a strong Feshbach resonance and strong inter-atomic collisions will therefore be of greater experimental utility, compared to those with a strong Feshbach resonance and weak inter-atomic collisions.

\begin{acknowledgments}
This work supported in part by the National Science Foundation (LK, HB, PP, MM, Grant Number 00900698) and the Undergraduate Research Program at Temple University (LK, HB).
\end{acknowledgments}


\begin{thebibliography}{99}
\bibitem{SPECTROSCOPY} The spectroscopy herein is different from free-bound-bound-ground transitions~\cite{KOK01} used in lattice experiments~\cite{NI08}, where a magnetic field couples atoms to Feshbach molecules, one laser couples these vibrationally-excited molecules to electronically-excited molecules, and another laser couples the electronically-excited molecules to meta-stable (triplet ground state) molecules.
\bibitem{SAT_NOTE} The saturation intensity, $I_0$, is defined~\cite{BOH99,MAC10,SCH02} as the intensity for which the stimulated decay rate equals the spontaneous decay rate, $\gamma_P(I_0)=\Gamma_s$.
\end{thebibliography}
\end{document}